\documentclass{ws-procs975x65}
\newcommand {\be}{\begin{equation}}
\newcommand {\ee}{\end{equation}}

\begin{document}

\title{A NEW PERSPECTIVE ON EARLY COSMOLOGY}

\author{EMANUELE ALESCI$^*$}

\address{Instytut Fizyki Teoretycznej, Uniwersytet Warszawski, ul. Ho{\.z}a 69, 00-681 Warszawa, Poland, EU\\
$^*$E-mail: emanuele.alesci@fuw.edu.pl}



\begin{abstract}
We present a new perspective on early cosmology based on Loop Quantum Gravity. We use projected spinnetworks, coherent states and spinfoam techniques, to implement a quantum reduction of the full Kinematical Hilbert space of LQG, suitable to describe inhomogeneous cosmological models. Some preliminary results on the solutions of the Scalar constraint of the reduced theory are also presented.
\end{abstract}


\bodymatter
\section*{}
Loop Quantum Gravity (LQG) \cite{revloop} describes a quantum space time at the Planck scale made of states of geometry called spinnetworks, eigenvectors of geometrical operators like area and volume with discrete spectra. This description of the Planck scale poses the problem of understanding how the classical spacetime emerges: 
there are good hints  for example from the graviton propagator \cite{ioprop} using the spinfoam dynamics \cite{eprl}, and from the semiclassical analysis using AQG \cite{cristina}. However the exact computations in the full theory are extremely difficult due to an heavy use of SU(2) recoupling theory and the absence of analytical closed formulas for the volume operator. A natural simplified arena to study the theory is then given by symmetry reduced models. The most efficient model in this sense is Loop Quantum Cosmology (LQC) \cite{lqc}. This is based on the implementation of a minisuperspace quantization scheme, in which the phase space is reduced on a classical level. Nevertheless  LQC is based on a ``loop inspired" quantization that leads to a polymer quantum mechanics while a direct derivation from LQG is still missing and it's difficult to accommodate in this setting inhomogenities because the theory is defined in the homogeneous reduced phase space.
If we try to relax the homogeneity assumption and we apply a loop quantization, we find a consistent algebra but   we loose\cite{fra, art} the graph structure proper of LQG and it's then not possible to apply Thiemann's procedure \cite{qsd} to define the Hamiltonian constraint.
Starting from these difficulties we introduce a new framework \cite{comp}  QRLG (Quantum Reduced Loop Gravity) to symmetry reduced models in which we start from the full LQG theory and implement at the quantum level the symmetry reduction. Here we apply QRLG to Bianchi I models \cite{comp}.

\paragraph{Quantum Reduced inhomogeneous Bianchi I}
The Bianchi I model describes a spatial manifold isomorphic to a 3-dimensional hyperplane with 1-forms $\omega^i=\delta^i_adx^a$.  We consider an inhomogeneous extension of this line element with each scale factor $a_i=a_i(t,x^i)$ function of time and of the corresponding Cartesian coordinate $x^i$ only. Within this assumption, viable in 1) the reparametrized Bianchi I model and 2) the generalized Kasner solution\cite{bkl} within a fixed Kasner epoch, we can infer the Ashtekar variables:
\be
A^i_a=c_i(t,x)\delta^i_a,\quad E^a_i=p^i(t,x)\delta^a_i.\label{in}
\ee
If we consider the vector dual to $\omega^i=\delta^i_a dx^a$ \emph{i.e} $\partial_i=\delta^a_i\partial_a$, the $SU(2)$ holonomies ${}^Rh^j_{e_i}$, along the edges $e_i$ parallel to $\partial_i$, associated with connections (\ref{in}) are given by
\begin{equation}
{}^Rh^j_{e_i}=P(e^{i\int_{e_i}c_idx^i(s)\tau_i}), \label{rhol}
\end{equation}
where $s$ is the arc length along $e_i$ and the latin indexes are not summed.
The fluxes across the surfaces $S^k$ dual to $e_k$, compatible with \eqref{in} are given by
\be
E_i(S^k)=\int E_i^a \delta^k_a dudv=\delta_i^k\int p_idudv.
\label{riduzione classica flussi}
\ee
The variables \eqref{in} can be derived by an SU(2) gauge-fixing that in terms of fluxes reads \cite{tu}
\be
\chi_i=\epsilon_{il}^{\phantom{12}k}E_k(S^l)=0.\label{gcon}
\ee
The conditions \eqref{rhol}, \eqref{riduzione classica flussi}, \emph{i.e} the holonomy-flux version of \eqref{in}, can be implemented in the kinematical Hilbert space of LQG i)by considering only edges $e_i$ parallel to fiducial vectors $\omega_i=\partial_i$ and ii) by the restriction from $SU(2)$ to $U(1)$ group elements via the imposition of the condition \eqref{gcon}. This is the quantum-reduction we consider.

The requirement i) can be realized via a projectors $P$ which acts on holonomies $h_e$ such that $Ph_e=h_e$ if $e=e_i$ for some $i$, otherwise it vanishes. Calling $\mathcal{H}_{P}$ the resulting Hilbert space,  we can project down to $\mathcal{H}_{P}$  the action of a generic diffeomorphisms and it turns out that only a subgroup of the whole 3-diffeomorphisms group is represented in $\mathcal{H}_P$ and we call its elements {\it reduced diffeomorphisms}.

As soon the condition (\ref{gcon}) is concerned, it cannot be implemented as first-class condition, because it is a gauge-fixing and it does not commute with the SU(2) Gauss constraint. Henceforth, we implement Eq. (\ref{gcon}) weakly, by mimicking the procedure adopted in Spin-Foam models to impose the simplicity constraints \cite{eprl}.

Imposing strongly the  Master constraint condition \cite{comp}, that arises extracting the gauge invariant part of $\chi_i$,
$
\chi^2=\sum_i\chi_i\chi_i=0
$
on $\mathcal{H_P}$, it will turn out that Eq. (\ref{gcon}) holds weakly and the classical relation (\ref{riduzione classica flussi}) can be implemented in a proper subspace of $\mathcal{H_P}$, as soon as $p_i$ are identified with the left invariant vector fields of the $U(1)_i$ groups generated by $\tau_i$. In fact if $\hat{\chi}^2$ is applied to a SU(2) $h^{j}_{e_i}$ one finds
\begin{equation}
\hat{\chi}^2 h^{j}_{e_i}=(8\pi\gamma l_P^2)^2(\tau^2-\tau_i\tau_i) h^{j}_{e_i}\;,
\label{classic condition}
\end{equation}
thus an appropriate solution to $\chi^2=0$ is given by
\be
 \tau^k h^{j}_{e_i} =0 \;,\quad \forall k\neq i.
 \label {solution weak} 
\ee

To find the quantum states that implement $\chi^2=0$ strongly and Eq. \eqref{solution weak} weakly, compatible with the restriction to holonomies of the kind \eqref{rhol}   we introduce particular functions over $SU(2)$, completely determined by their restriction to the $U(1)_i$ subgroups generated by $\tau_i$, using the Dupuis-Livine map \cite{dl}. These functions are 
\be
\tilde{\psi}(g)_{e_i}=\sum_{j} {}^i\!D^{|j|}_{jj}(g) \psi^{j}_{e_i}\quad,\; \quad g\in SU(2)
\label{fine projected}
\ee
with basis elements solutions of the Master constraint $
{}^i\!D^{j}_{jj}(g)= \langle j,\vec{e_i}|D^j(g)|j,\vec{e}_i\rangle
$
for $i=1,2,3$ with $|j,\vec{e}_i\rangle$ standard $SU(2)$ coherent states in the direction $\vec{e}_i$ tangent to $e_i$ and $D^j(g)$ Wigner matrices.
These states also satisfy the condition \eqref{solution weak} weakly:
\be
\langle\tilde\psi'_i|\hat{E}_k(S^l)|\tilde\psi_i\rangle=8\pi \gamma l_P^2 \sum_{j,j'}\psi^{j'}_{e_i}\int dg  {}^i\!D^{j'}_{j'j'}(g)\tau_k {}^i\!D^j_{jj}(g)\psi^{j}_{e_i}=0,\qquad k\neq i.\label{weak}
\ee
In this way the resulting quantum states associated with an edge $e_i$ are entirely determined by their projection into the subspace with maximum (minimum) magnetic numbers along the internal direction $i$. 
We call the $SU(2)$ states of the form \eqref{fine projected} \emph{ quantum-reduced states} and they define a subspace of $\mathcal{H}_{P}$ denoted $\mathcal{H}^{R}$.
However, because we have established a projection from $\mathcal{H}^{kin}$ to $\mathcal{H}^R$, we can implement the original Gauss, Diff and Hamiltonian constraints defined on  $\mathcal{H}^{kin}$ by projection on $\mathcal{H}^R$ . The Gauss constraint is solved introducing some {\it reduced intertwiners}, (the standard $SU(2)$ ones projected on  $\mathcal{H}^R$), the Diff constraint is solved by considering reduced s-knot states (s-knots restricted to reduced diffeomorphisms) and the Hamiltonian constraint can be explicitly computed\cite{art} because the volume is diagonal on $\mathcal{H}^R$. We have then an anomaly free theory (the algebra of the constraint works like in full LQG) that behaves as a $U(1)$ theory along the edges but keeps an $SU(2)$ structure at the nodes allowing to take into account anisotropies and inohomogenieties, in an exactly computable framework\cite{art}.

\paragraph{Acknowledgments} The work of the author was partially supported by the grant of Polish Narodowe Centrum Nauki nr 2011/02/A/ST2/00300.

\end{document}